\documentclass{optica-article}

\journal{opticajournal} 

\articletype{Research Article}

\usepackage{graphicx,float}
\usepackage{subfigure}
\usepackage{dcolumn}
\usepackage{bm}
\usepackage{CJK}
\usepackage{mathrsfs}
\usepackage{amsmath}
\usepackage{siunitx}
\newcommand{\ket}[1]{|#1\rangle}
\newcommand{\bra}[1]{\langle #1|}
\usepackage{lipsum}
\usepackage{color}
\usepackage{cite}
\usepackage{placeins}

\begin{document}

\title{Generation of True Quantum Random Numbers with On-Demand Probability Distributions via Single-Photon Quantum Walks}

\author{Chaoying Meng,\authormark{1,3,$\dag$} Miao Cai, \authormark{2,3,$\dag$} Yufang Yang, \authormark{1,3} Haodong Wu, \authormark{2,3} Zhixiang Li, \authormark{2,3} Yaping Ruan, \authormark{2,3} Yong Zhang, \authormark{2} Han Zhang, \authormark{1,3,*} Keyu Xia, \authormark{2,3,4*} Franco Nori,\authormark{5,6}}

\address{\authormark{1}School of Physics, Nanjing University, Nanjing 210023, China\\
\authormark{2}College of Engineering and Applied Sciences, and National Laboratory of Solid State Microstructures, Nanjing University, Nanjing 210023, China\\
\authormark{3}National Laboratory of Solid State Microstructures, Nanjing University, Nanjing 210093, China\\
\authormark{4}Shishan Laboratory, Suzhou Campus of Nanjing University, Suzhou 215000, China\\
\authormark{5}Quantum Computing Center, Cluster for Pioneering Research, RIKEN, Wakoshi, Saitama 351-0198, Japan\\
\authormark{6}Physics Department, The University of Michigan, Ann Arbor, Michigan 48109-1040, USA\\
\authormark{$\dag$}These two authors contributed equally}

\email{\authormark{*}zhanghan@nju.edu.cn; keyu.xia@nju.edu.cn} 

\begin{abstract*} 
Random numbers are at the heart of diverse fields, ranging from simulations of stochastic processes to classical and quantum cryptography. The requirement for true randomness in these applications has motivated various proposals for generating random numbers based on the inherent randomness of quantum systems. The generation of true random numbers with arbitrarily defined probability distributions is highly desirable for applications, but it is very challenging. Here we show that single-photon quantum walks can generate multi-bit random numbers with on-demand probability distributions, when the required ``coin'' parameters are found with the gradient descent (GD) algorithm. Our theoretical and experimental results exhibit high fidelity for various selected distributions. This GD-enhanced single-photon system provides a convenient way for building flexible and reliable quantum random number generators. Multi-bit random numbers are a necessary resource for high-dimensional quantum key distribution.

\end{abstract*}

\section{Introduction}\label{sec:intro}
Random numbers are important for science research and engineering applications, such as Monte-Carlo simulations~\cite{Nicholas1949,Niederreiter1978}, cryptography~\cite{Shannon1949,Gennaro2006} and tests of fundamental physics~\cite{Bell1964,Shadbolt2014}. For example, quantum key distribution (QKD) technology highly relies on the availability of true random numbers to protect its communication security~\cite{Gisin2002,Scarani2009,Lo2014,Martin2015}. Theoretically, pseudo-random number generators, due to their deterministic and predictable nature, cannot satisfy the requirement for building perfectly secure communication systems. Therefore, the inherent randomness of a quantum system makes it a promising platform for generating faithful random numbers. This is known as quantum random number generator (RNG)\cite{Herrero2017}.

Practical quantum RNGs using various sources of randomness have been demonstrated. Discrete generators can use branching paths\cite{Rarity1994,Jennewein2000,Andre2000}, arrival times\cite{Ma2005,Stipcevic2007,Dynes2008,Michael2009}, photon counting\cite{Harald2010,Ren2011,Jian2011,Tisa2015}, and attenuated pulse\cite{Wei2009,Zahra2015}; whereas continuous approaches exploit quantum vacuum fluctuations\cite{Shen2010,Gabriel2010,Zhu2012}, phase noise of lasers\cite{Guo2010,Qi2010,Nie2015}, amplified spontaneous emission\cite{Martin2015,Williams2010}, and Raman scattering\cite{Bustard2011}. Among these schemes, quantum RNG based on quantum walks promise a convenient and fast way to generate true random numbers\cite{Sarkar2019}.

The applications of a RNG strongly rely on the probability distribution used. Different distributions are indispensable in various fields. Uniformly distributed random numbers are most desirable and particularly useful in practical applications~\cite{Herrero2017} because these avoid inherent bias. A Gaussian distributed RNG is of most significance in the modulation of coherent states in continuous-variable QKD systems~\cite{Zhang2019a,Zhang2020,Huang2020}, simulations of communication channels, and stochastic processes (e.g. noise)~\cite{Thomas2007}. 

It is highly valuable to develop a quantum RNG with an on-demand probability distribution. Based on quantum walks, significant efforts have been made for this task~\cite{Sarkar2019,Zhang2022}. However, it is challenging to find the proper parameter numbers for a complex system to generate true random numbers with a given distribution. In contrast, the gradient descent (GD) algorithm, as a highly adaptive optimization algorithm that has been widely utilized in many fields~\cite{Biamonte2017, Pantita2017, Kerenidis2020, Cai2021}, can provide a more general and efficient way to accomplish this challenging task.

In this work, we propose a GD-enhanced quantum walk for realizing quantum RNG with, in principle, an on-demand probability distribution. Our GD-based scheme can be implemented by using a linear optical system without the need of time-bin encoding and dynamical modulation. We further experimentally demonstrate the generation of true random numbers with various selected probability distributions by using quantum walks of heralded single photons.  

\section{System and model}\label{sec:sysandmod}
In quantum walks, the walker is located in the Hilbert space $\mathcal{H}\equiv\mathcal{H}_{\text{p}}\otimes\mathcal{H}_{\text{c}}$, where $\mathcal{H}_{\text{p}}$ is position space and $\mathcal{H}_{\text{c}}$ is the coin space. The coin space contains two basis vectors $\{\ket{L},\ket{R}\}$, which represent the eigenstate of the coin. Therefore, the definite position and classical coins are both replaced by position states and coin operators in a quantum walk system. 

In a one-dimensional (1D) discrete-time quantum walk system, the quantum walker's state can be described by a product state $\ket{\Psi}=\ket{\psi}\otimes\ket{c}$, where $\ket{c}=\alpha_{L}\ket{L}+\alpha_{R}\ket{R}$ is the coin state and $\ket{\psi}=\sum_{x}{\alpha_{x}\ket{x}}$ is the position state. Each walking step consists of a unitary operator $\hat{U}=\hat{S}\hat{C}$, where $\hat{S}$ is the conditional shift operator and $\hat{C}$ is the coin operator. The coin operator $\hat{C}$ rotates the coin state and its most general form can be expressed as 
\begin{equation}\label{eq:Eq1}
\hat{C}=\sum_{x}{\ket{x}\bra{x}}\otimes \text{e}^{i\beta}\left(  
\begin{matrix}
\text{e}^{i\xi} \cos(\theta) & \text{e}^{i\zeta} \sin(\theta) \\
-\text{e}^{i\zeta} \sin(\theta) & \text{e}^{-i\xi} \cos(\theta)
\end{matrix}
\right) \; ,
\end{equation}
where $\xi ,\zeta \in [0,2\pi]$ and $\theta\in[0,\pi/2]$ are the parameters of the rotation and $\beta$ fixes the global phase. The conditional shift operator $\hat{S}$ moves the walker either to the left or right depending on the coin state and has the form
\begin{equation}\label{eq:Eq2}
\hat{S} = \sum_{x}{\ket{x-1,L}\bra{x,R}+\ket{x+1,R}\bra{x,L}}\; .
\end{equation}
It leads to the conditional shift operation $\hat{S}\ket{x,L} = \ket{x+1,L}$ and $\hat{S}\ket{x,R} = \ket{x-1,R}$. In the following, we fix the parameters $\beta = \pi/2$ and $\xi = \zeta = -\pi/2$, so that we obtain the coin determined by one parameter $\theta$. If $\theta=\pi/4$, the coin then becomes the Hadamard coin : $\hat{H} = \left[\begin{matrix} 
1 & 1 \\
1 & -1 
\end{matrix}\right]/\sqrt{2}$.

After $n$ walking steps, the state of a quantum walk system becomes $\ket{\Psi_{n}}$. The quantum walker remains in a superposition of many positions until the final measurement is performed. The measured probability for the walker being at $x_{k}$ after $n$ walking steps can be written as
\begin{equation}\label{eq:Eq3}
\mathcal{P}(x_{k})=|\bra{R}\langle x_{k}\ket{\Psi_{n}}|^2 + |\bra{L}\langle x_{k}\ket{\Psi_{n}}|^2\; .
\end{equation}
The probability distribution is determined by the choice of coin parameter set in each walking step. It is difficult to adjust the coin parameters to obtain desired probability distributions because the number of coin parameters grows rapidly when increasing the walking steps. In this work, we exploit the gradient descent algorithm to solve this challenging problem.

\section{Algorithm}\label{sec:algo}
Generally, a GD algorithm consists of three elements~\cite{Ruder2016}: a system function $F$, system parameters $\{\theta_{i}\}\;(i=1,2,...,k)$, and a loss function $\mathcal{L}$. The system function $F$ defines the input-output relation of the system and is parameterized by the parameters $\{\theta_{i}\}\;(i=1,2,...,k)$. The loss function $\mathcal{L}$ evaluates the system output and compares it to the target. Here, we write the system function $F$ as $y = F(x\; ;\theta_{1},\theta_{2},...,\theta_{k})$, where $x$ and $y$ are the input and output of the system, respectively. Hereafter we use the mean square error function $\mathcal{L}=\frac{1}{2}(T-y)^2$ as the loss function, where $T$ is the target. Therefore, the loss function $\mathcal{L}$ is also parameterized by $\{\theta_{i}\}\;(i=1,2,...,k)$ and can be written as $\mathcal{L}(\{\theta_i\})$. Essentially, the gradient descent method minimizes the loss function $\mathcal{L}(\{\theta_i\})$ by updating $\{\theta_{i}\}$ in the opposite direction of the gradient of the loss function. In each iteration of the gradient descent method, the parameters $\theta_{i}$ are updated according to $[\theta_{i}-\eta\cdot\nabla_{\theta_{i}}\mathcal{L}(\theta_{i})] \rightarrow \theta_{i}$\cite{Ruder2016}, where $\eta\in(0,1]$ is the learning rate.

\begin{figure}
  \centering
  \includegraphics[width=1.0\linewidth]{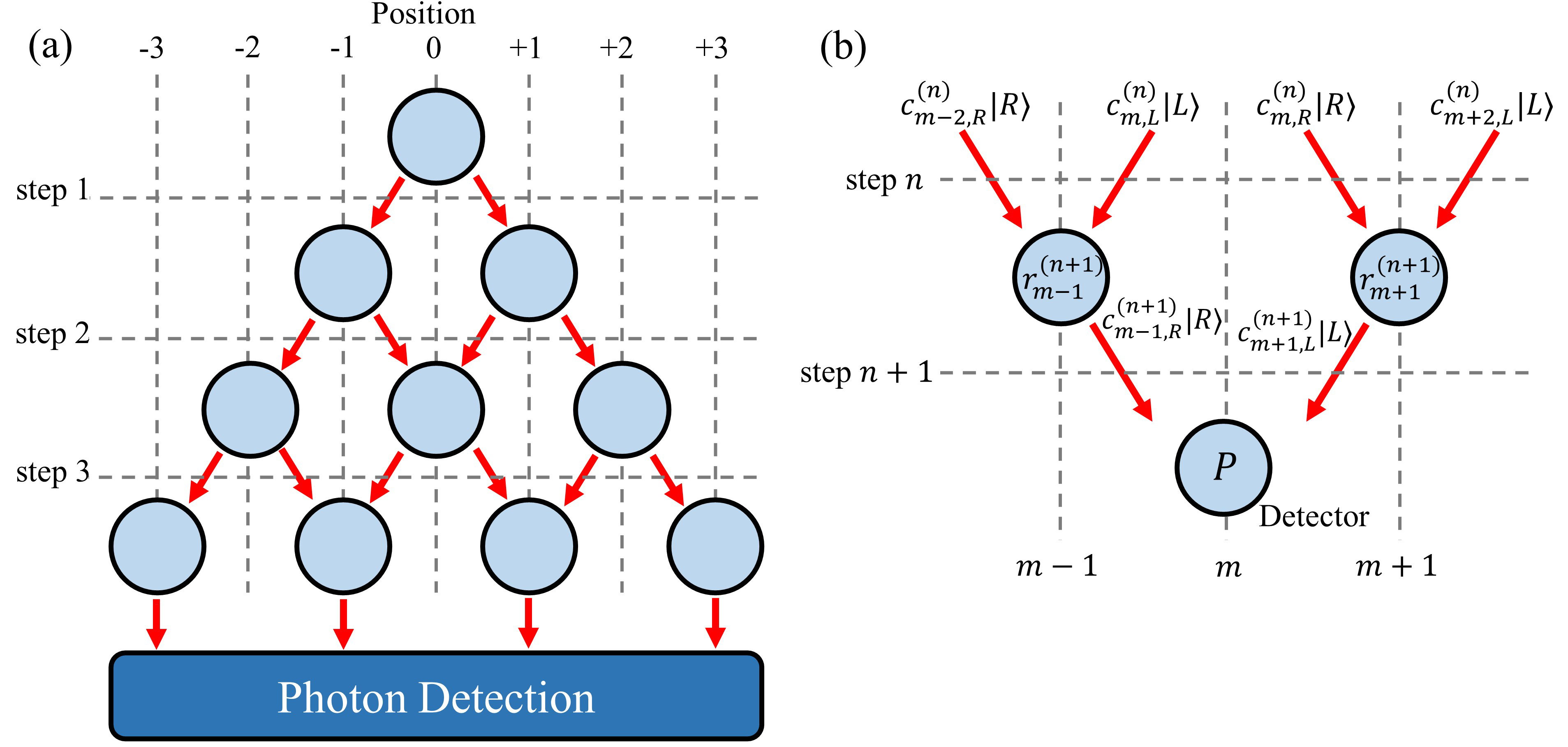} \\
\caption{(a) Schematic of a one-dimensional discrete-time quantum walk process. The red arrows represent the walking directions. The vertical and horizontal gray dashed lines denote the position states and the walking steps, respectively. (b) Details of the quantum state transfer in a quantum walk. The symbols next to the red arrows describe the coin state transfer in each walking step. The symbol $r_{m}^{n}$ represents the coin bias ratio of the $n$-th walking step starting from position $m$.}
\label{fig:FIG1}
\end{figure}

A 1D discrete-time quantum walk process is depicted in Fig.~\ref{fig:FIG1} (a). The blue circles denote different position states, and the red arrows indicate the directions of the walk starting from different position states in each walking step. Without loss of generality, we assume the coin bias ratio can be adjusted for every coin operation at different position states in different walking steps. This assumption can be experimentally realized in a linear optical system~\cite{Jeong2004}.

The specific description of quantum state transfer in a quantum walk system is shown in Fig.~\ref{fig:FIG1}(b). The letters $c$ and $r$ represent the complex amplitude and coin bias ratio, respectively. The notation $c_{m,R}^{(n)}\;$($c_{m,L}^{n}$) represents the complex amplitude of the coin state $\ket{R}$ ($\ket{L}$) at the position $m$ in the $n$-th walking step; while $r_{m}^{(n)}$ is the coin bias ratio of the $n$-th walking step starting from position $m$, and $P_{m}$ is the measured probability at the detector located at position $m$. The coin bias ratio $r$ is defined as $r = \cos^2\theta$. Thus, the state transformation with coin bias ratio $r$ can be modeled as $c\ket{L}\rightarrow\sqrt{r}\cdot c\ket{L} + \sqrt{1-r}\cdot c\ket{R}$, and $c\ket{R}\rightarrow\sqrt{1-r}\cdot c\ket{L} - \sqrt{r}\cdot c\ket{R}$. Then the measured probability $P_{m}$ becomes
\begin{equation}\label{eq:Eq4}
\begin{split}
P_{m}  = &  \left[ \sqrt{1 - r_{m-1}^{(n+1)}}a_{m-2,R}^{(n)} - \sqrt{r_{m-1}^{(n+1)}}a_{m,L}^{(n)}\right]^2 + \left[\sqrt{1 - r_{m-1}^{(n+1)}}b_{m-2,R}^{(n)} - \sqrt{r_{m-1}^{(n+1)}}b_{m,L}^{(n)}\right]^2 \\
& + \left[\sqrt{r_{m+1}^{(n+1)}}a_{m,R}^{(n)} + \sqrt{1 - r_{m+1}^{(n+1)}}a_{m+2,L}^{(n)}\right]^2 + \left[\sqrt{r_{m+1}^{(n+1)}}b_{m,R}^{(n)} + \sqrt{1 - r_{m+1}^{(n+1)}}b_{m+2,L}^{(n)}\right]^2\; ,
\end{split}
\end{equation}
where $a$ and $b$ are the real and imaginary components of $c$, respectively.

According to the GD algorithm, the updated value of $r_{m}^{(n)}$ with respect to $P_{j}$ is
\begin{equation}\label{eq:Eq5}
\Delta r_{m,P_{j}}^{(n)} = -\eta\frac{\partial \mathcal{L}}{\partial r_{m}^{(n)}} = -\eta\frac{\partial \mathcal{L}}{\partial P_{j}}\frac{\partial P_{j}}{\partial r_{m}^{(n)}}=\eta(T_{j}-P_{j})\frac{\partial P_{j}}{\partial r_{m}^{(n)}}\; ,
\end{equation}
where here the loss function becomes $\mathcal{L} = \frac{1}{2}\sum_{j}(T_{j}-P_{j})^2$, and $T_{j}$ is the target probability at position $j$. Then the overall updated value of $r_{m}^{(n)}$ is obtained by summing Eq.~(\ref{eq:Eq5}), 

\begin{equation}\label{eq:Eq6}
\sum_{j}\Delta r_{m,P_{j}}^{(n)}=\sum_{j}\eta(T_{j}-P_{j})\frac{\partial P_{j}}{\partial r_{m}^{(n)}}\; .
\end{equation}

The details of the derivation are presented in the supplemental document. Therefore, during each iteration of our algorithm, $r_{m}^{(n)}$ updates according to the following relation
\begin{equation}\label{eq:Eq7}
\left[r_{m}^{(n)}+\sum_{j}\eta(T_{j}-P_{j})\frac{\partial P_{j}}{\partial r_{m}^{(n)}}\right] \rightarrow r_{m}^{(n)} \; .
\end{equation}

The training finishes when the simulated quantum walk probability distribution reaches the target distribution. After the training is completed, the theoretical values of the coin bias ratios for generating the desired probability distribution are obtained.

\section{Experimental setup}\label{sec:expset}
Quantum walks lay the natural foundation for studying plenty of novel quantum phenomena and can be realized in various systems\cite{Perets2008, Alberto2010, Tang2018, Su2019, Yan2019, Su2022}. Among these, linear-optics-based quantum walks have advantages in convenience of implementation and compatibility. Therefore, we use this platform for our GD-based quantum RNG scheme.

In linear optical implementations of quantum walks, we use single photons as the quantum walker that moves in both directions in every position. The polarization states $\{\ket{H},\ket{V}\}$ are introduced to represent two orthogonal coin states $\{\ket{L},\ket{R}\}$, respectively. We use single-photon spatial modes to represent the position of the walker $\ket{x}$.

\begin{figure}
  \centering
  \includegraphics[width=1\linewidth]{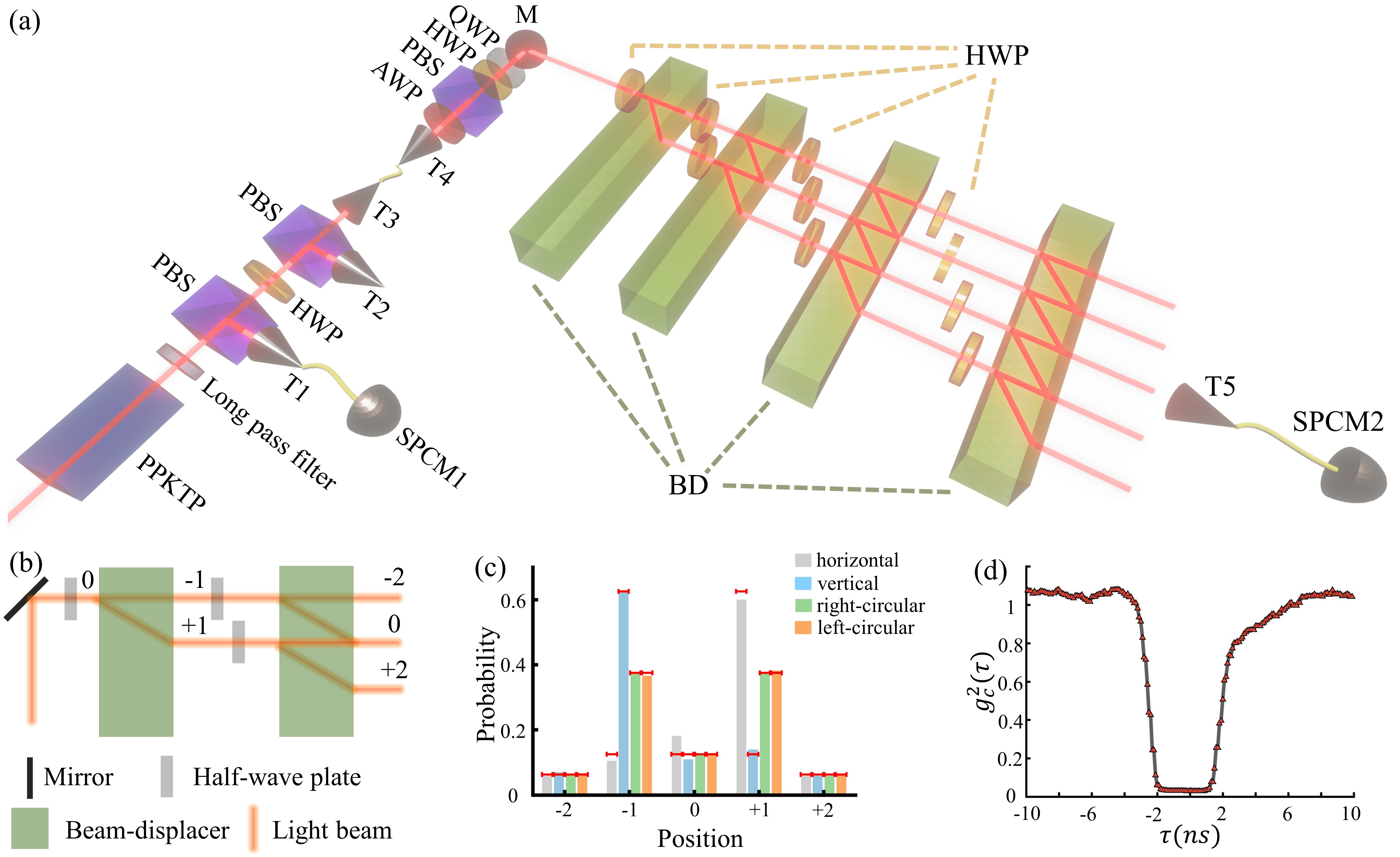} \\
\caption{(a) Schematic of experimental setup. PPKTP: periodically poled potassium titanyl phosphate crystal, PBS: polarized beam splitter, SPCM: single photon counting module, HWP: half-wave plate, QWP: quarter-wave plate, AWP: adjustable wave plate, M: mirror, BD: beam displacer. Here AWP is designed as a HWP in the middle of two QWPs in order to compensate for the phase shift caused by the fiber twist, and can convert circular polarized light to horizontal polarized light without loss. (b) Details of the first two quantum walk steps in our experiment. (c) Measured (colorful bars) and theoretical (red dashes) probability distribution for a four-step quantum walk. Different colors represent different polarization states of the initial input single photons. (d) The second-order correlation function $g_{c}^2(\tau)$ versus the delay $\tau$ for our single-photon source. The time window length is approximately $3$~\unit{\nano\second}~and $g_{c}^{2}(0)$ is $0.0286\pm 0.001$.}
\label{fig:FIG2}
\end{figure}

The schematic of our experimental setup is shown in Fig.~\ref{fig:FIG2}(a). Pairs of single photons are created via type-II spontaneous parametric down-conversion in a periodically poled potassium titanyl phosphate (PPKTP) crystal. This crystal is pumped by a diode laser centered at $397.5$~\unit{\milli\meter}~and emits orthogonally polarized photon pairs (i.e., horizontal and vertical polarized, or left- and right-circularly polarized) with a wavelength of 795~\unit{\nano\meter}~and a FWHM bandwidth of 0.3~\unit{\nano\meter}. The photon pairs are separated by a polarized beam splitter. One photon from each pair served as a trigger while the other photon is launched into the quantum walk system. 

The position states of the quantum walk are represented by spatial modes of the single photons. The shift operator $\hat{S}$ acting on these modes is implemented by a 37.7~\unit{\milli\meter}~long, birefringent calcite beam displacer. The optical axis of each calcite prism is cut so that vertically polarized light was directly transmitted, and horizontal light underwent a 4~\unit{\milli\meter}~lateral displacement into a neighboring spatial mode. Here, we place the half wave plates in front of each beam displacer to adjust the coin bias ratio in the quantum walk. The aperture diameter of our half wave plate is small so that each half wave plate can change the polarization state of one beam of light without affecting adjacent beams. Therefore, we can adjust the coin bias ratio at different positions during each walking step.

The details of the first two quantum walk steps are depicted in Fig.~\ref{fig:FIG2}(b). The spatial modes after step $1$ are recombined interferometrically in step $2$. Repetition of these steps then forms an interferometric network as in Fig.~\ref{fig:FIG2}(a). The lattice sites are labeled so that there are odd sites at odd walking steps, and even sites at even steps. After an $n$-step quantum walk, the photons output in ($n+1$) spatial modes are coupled into an optical fiber and subsequently detected by a single-photon photodiode, in coincidence with the trigger photon. We measure the final probability distribution of the quantum walk by manually moving the fiber coupler between individual output spatial modes.

For a four-step quantum walk with an unbiased coin ($\theta = \pi/4$), the measured probability distribution at given sites is shown in Fig.~\ref{fig:FIG2}(c). Here we choose four initial polarization states to verify our experimental system: horizontal polarization, vertical polarization, right-circular polarization, and left-circular polarization. The experimental data (bars with colors) are in excellent agreement with theoretical simulations (red dashes). To characterize the single-photon purity in the experiment, we also measure the second-order correlation function $g_{c}^{2}(\tau)$ for our single-photon source, as depicted in Fig.~\ref{fig:FIG2}(d).

\section{Results}\label{sec:result}
\subsection{Uniform distribution}
Quantum RNGs with a uniform distribution~\cite{Ren2011, Eaton2022} are of importance for applications without inherent bias, such as quantum secure communication~\cite{Tang2014, Herrero2017}. Therefore, we first evaluate the performance of our algorithm for generating a uniform distribution in a four-step quantum walk system. Here we use the fidelity $\mathcal{F}$, defined to evaluate the similarity between the output (simulated or measured output) and the target probability distribution,
\begin{equation}\label{eq:Eq8}
\mathcal{F} = \frac{\sum_{m}{y(m)\cdot T(m)}}{\sum_{m}\text{max}(y(m), T(m))^2}\; ,
\end{equation}
where $y$ is the system output, $T$ is the target distribution, and $m$ represents the position.

\begin{figure}
  \centering
  \includegraphics[width=0.6\linewidth]{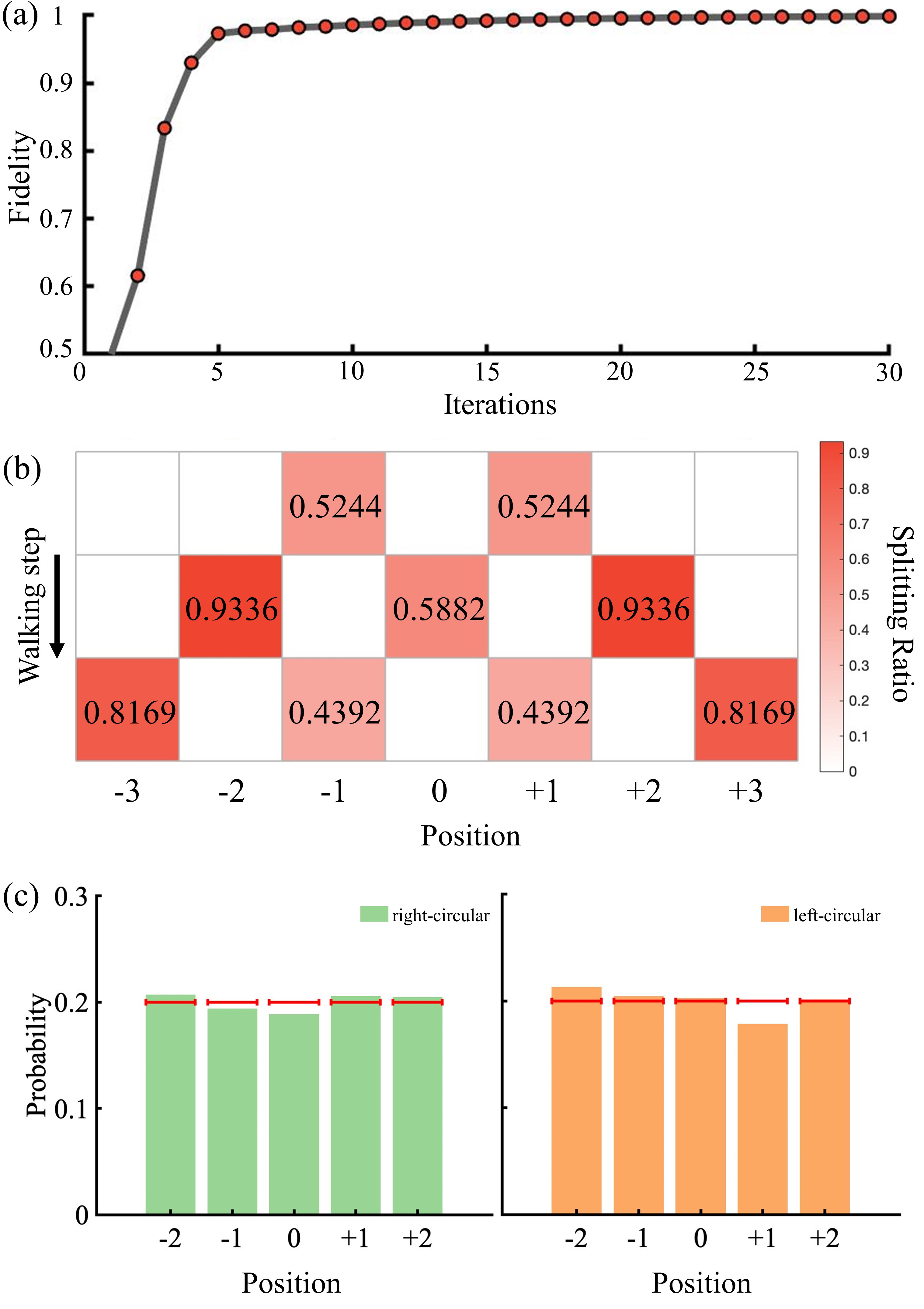} \\
\caption{Uniform probability distribution generation in a four-step quantum walk system. (a) Fidelity as the iteration increases. (b) Values of the coin bias ratio of each position and walking step, obtained with our GD algorithm. The black arrow points out the direction of the quantum walking process. The number displayed on the cell is the value of the corresponding coin bias ratio $r$. (c) Measured probability distribution of the quantum walk for right-circular (green) and left-polarized (orange) single photons. The red dashes represent the values of the target probability distribution.}
\label{fig:FIG3}
\end{figure}

For generating a uniform probability distribution, the fidelity curve during the training of our GD algorithm is shown in Fig.~\ref{fig:FIG3}(a). The ``iterations" represent the accumulating time step when the training progresses. From Fig.~\ref{fig:FIG3}(a) we can see that the fidelity increases rapidly as the training process goes on. It exceeds $0.95$ after $5$ iterations and finally approaches unity within $20$ iterations. The learning rate of the training process is set as $0.1$. The convergence rate of the training can be further improved by appropriately choosing the learning rate $\eta$. 

When the training is completed, we obtain the values of the coin bias ratio for generating a uniform probability distribution in the quantum walk. The values are shown in Fig.~\ref{fig:FIG3}(b). Obviously, these values are unlikely to be found manually, while our algorithm can find proper values to obtain a high fidelity. According to these values, we adjust $\{r\}$ in the quantum walk experimental setup by rotating the half-wave plates in front of the BDs. The minimal adjustable angle of our half-wave plate is $0.25$ degree, which leads to a slight deviation between the actual coin bias ratio in the experiment and the theoretical values. But this does not affect the performance of our experiment because our system has strong robustness (See supplemental document for details of the experimental system robustness analysis). We perform experiments with right-circular and left-circular polarized single photons, respectively. The measured probability distributions for detecting the photon at given positions are shown in Fig.~\ref{fig:FIG3}(c). It is clear that the measured probability distributions are in good agreement with the target distribution. The fidelities of the experimental results are $96.5\%$ for right-circular polarized photons and $95.8\%$ for left-circular polarized photons.

\begin{figure}
  \centering
  \includegraphics[width=0.6\linewidth]{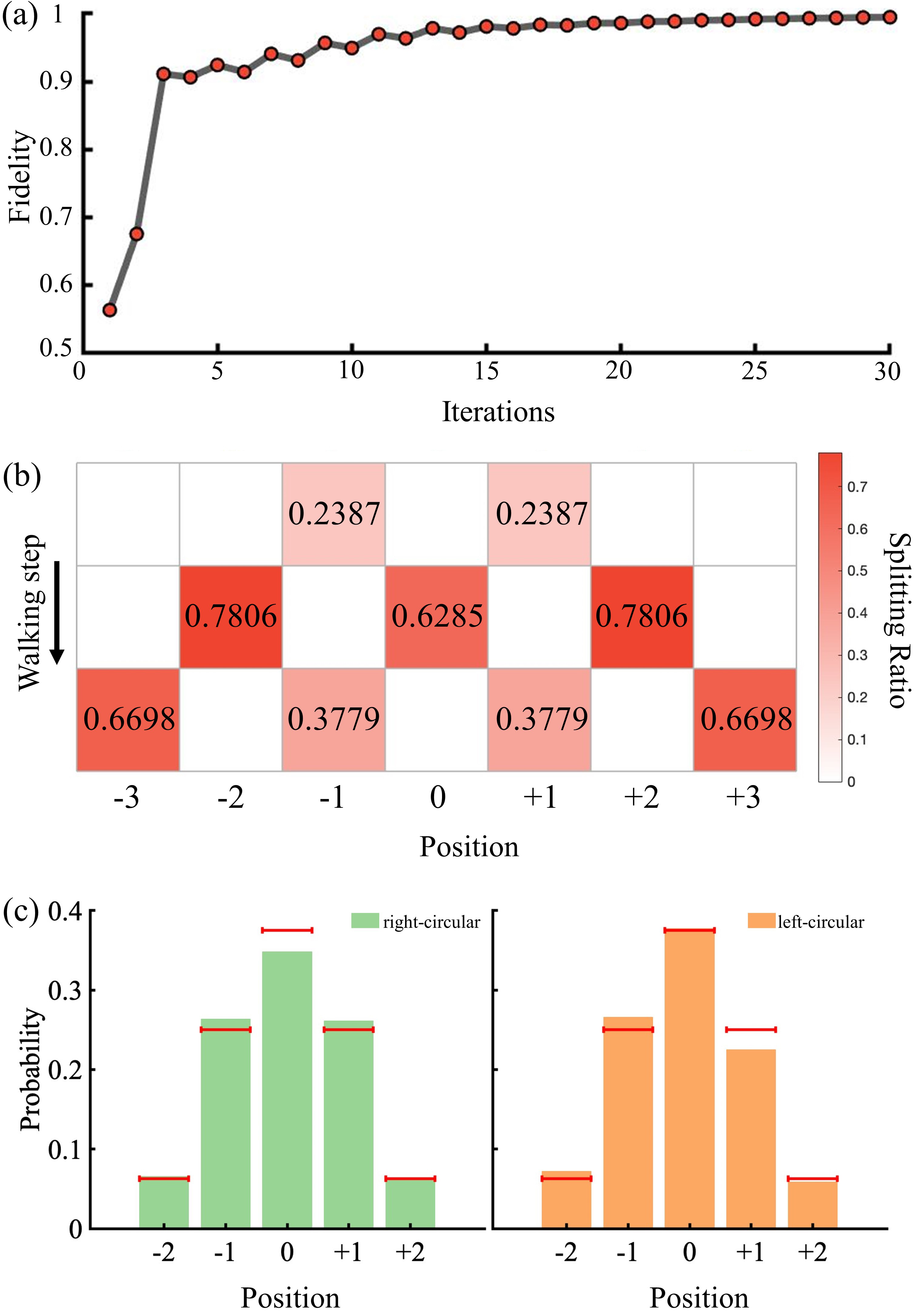} \\
\caption{Gaussian probability distribution generation in a four-step quantum walk system. (a) Fidelity as the iteration increases. (b) Values of the coin bias ratio for each walking step and position, obtained with our GD algorithm. (c) Measured probability distribution of the quantum walk for right-circular (green) and left-polarized (orange) single photons.}
\label{fig:FIG4}
\end{figure}

\subsection{Gaussian distribution}
Gaussian RNGs, as another important RNG, also have diverse useful applications, including Monte Carlo simulation of Gaussian noises. Specific to quantum information, this type of RNGs provide Gaussian distributed randomness for coherent states modulation in continuous-variable quantum key distribution systems~\cite{Zhang2019a,Zhang2020,Huang2020}. In the following, we show that our GD algorithm can find the parameter set for the quantum walk based RNG to generate Gaussian distributed single-photon outputs.

We set the Gaussian distribution as the target probability distribution for the GD algorithm. The fidelity change during the training process is shown in Fig.~\ref{fig:FIG4}(a). It can be seen that the fidelity rapidly increases to $95\%$ at the $10$th iteration. The coin bias ratios can be found in Fig.~\ref{fig:FIG4}(b). Figure~\ref{fig:FIG4}(c) presents the measured probability distribution of single photons in a quantum walk with GD-optimized coin bias ratios. Right- and left-circularly polarized photons are chosen as input photons to perform the quantum walk experiment. The experimentally measured probability distribution is again in good agreement with the target distribution. The fidelities of the experiment results are $94.1\%$ and $95.8\%$ for the right- and left-circular polarized input photons, respectively. These results show that our algorithm can be utilized to adjust a quantum walk system to generate single photons with desired distributions. This allows one to build an effective quantum RNG that conforms to arbitrary probability distributions.

\section{Conclusion}\label{sec:conc}
We have reported a GD-enhanced quantum RNG based on quantum walks of single photons in a linear optical system. Our multi-bit quantum RNG can generate true random numbers with an arbitrarily defined probability distribution with nearly unitary fidelity. The promised faithful randomness of our quantum RNG can determine the random measurement basis in high-dimensional quantum communications~\cite{Vaziri2003, Wang2018, Liu2022, Li2022}. We note that quantum walks with a uniform distribution can be used to generate quantum random numbers~\cite{Grafe2014}. In comparison with this method, our GD-enhanced quantum walk can generate quantum random numbers with flexible probability distribution.

\begin{backmatter}
\bmsection{Funding}
National Key R\&D Program of China (Grant No. 2019YFA0308700), the National Natural Science Foundation of China (Grant No.11890704), the Program for Innovative Talents and Teams in Jiangsu (Grant No. JSSCTD202138), Nippon Telegraph and Telephone Corporation (NTT) Research, the Japan Science and Technology Agency (JST) (via the Quantum Leap Flagship Program (Q-LEAP), and the Moonshot R\&D Grant Number JPMJMS2061), the Asian Office of Aerospace Research and Development (AOARD) (Grant No. FA2386-20-1-4069), and the Foundational Questions Institute Fund (FQXi) (Grant No. FQXi-IAF19-06).

\bmsection{Acknowledgements}
The authors thank Lijian Zhang and Ben Wang for helpful discussions. We thank the High Performance Computing Center of Nanjing University for allowing the numerical calculations on its blade cluster system.

\bmsection{Disclosures}
The authors declare that there are no conflicts of interest related to this article.

\bmsection{Supplemental document}
See Supplement 1 for supporting content. 

\end{backmatter}



\end{document}